\documentclass[aip,reprint,superscriptaddress,floatfix]{revtex4-1}

\usepackage{mathptmx}
\usepackage[scaled=0.92]{helvet}
\usepackage[utf8]{inputenc}
\usepackage{graphicx}
\usepackage[dvipsnames]{xcolor}
\usepackage[pdftex]{hyperref}
\hypersetup{colorlinks,citecolor=blue}

\usepackage{newunicodechar}
\newunicodechar{⁠}{}

\usepackage[textwidth=1.5cm]{todonotes}
\usepackage{textcomp}
\usepackage{ifsym}
\usepackage{mathrsfs}
\usepackage{sidecap}
\usepackage[
,textwidth=17.5cm
,textheight=23.5cm
,verbose
,dvips
]{geometry}
\usepackage{xspace}

\newcommand{\celsius}{$^{\circ}$C\xspace}

\newcommand{\ts}{\(T_{\mathrm{S}}\)\xspace}
\newcommand{\fal}{\(f_{\mathrm{Al}}\)\xspace}
\newcommand{\fn}{\(f_{\mathrm{N}}\)\xspace}
\newcommand{\rsh}{\(R_{\mathrm{sh}}\)\xspace}

\begin{document}

\title{Growth-induced trade-off between polarity, electrical isolation, and structural quality in AlN grown on Si(111) by plasma-assisted molecular beam epitaxy}

\author{Hossein~Yazdani}
\email{yazdani@pdi-berlin.de}
\affiliation{Paul-Drude-Institut für Festkörperelektronik (PDI), Leibniz-Institut im Forschungsverbund Berlin e.V., 10117 Berlin, Germany}
\author{Nhu~Tran}
\affiliation{Paul-Drude-Institut für Festkörperelektronik (PDI), Leibniz-Institut im Forschungsverbund Berlin e.V., 10117 Berlin, Germany}
\affiliation{Institute of Physics of Materials and CEITEC IPM, Czech Academy of Sciences, 616~00 Brno, Czech Republic}
\author{Xin~Du}
\affiliation{Paul-Drude-Institut für Festkörperelektronik (PDI), Leibniz-Institut im Forschungsverbund Berlin e.V., 10117 Berlin, Germany}
\author{Vit~Hornik}
\affiliation{Institute of Physics of Materials and CEITEC IPM, Czech Academy of Sciences, 616~00 Brno, Czech Republic}
\author{Roman~Gröger}
\affiliation{Institute of Physics of Materials and CEITEC IPM, Czech Academy of Sciences, 616~00 Brno, Czech Republic}
\author{YongJin~Cho}
\email{cho@pdi-berlin.de}
\affiliation{Paul-Drude-Institut für Festkörperelektronik (PDI), Leibniz-Institut im Forschungsverbund Berlin e.V., 10117 Berlin, Germany}
\affiliation{School of Electrical and Computer Engineering, Cornell University, Ithaca, New York 14853, USA}

\begin{abstract}
We investigate the effects of substrate temperature and Al flux, at a fixed N flux, on the polarity, electrical properties, and structural quality of AlN films grown on high-resistivity Si(111) by plasma-assisted molecular beam epitaxy. N-rich conditions result in relatively highly resistive Al-polar AlN/Si, whereas Al-rich conditions promote N-polarity and improved structural quality but simultaneously lead to increased \textit{p}-type conductivity of AlN/Si stacks, consistent with unintentional Al doping of the Si substrate during AlN growth. Increasing the substrate temperature further improves the structural quality while enhancing the unwanted conductivity, revealing a growth-induced trade-off among polarity, electrical isolation, and structural quality. By reducing the substrate temperature to 575\celsius while maintaining slightly Al-rich conditions, an N-polar AlN/Si stack with a sheet resistance of 276~k$\Omega$/sq is achieved. These results provide guidelines for optimizing AlN/Si templates for various nitride device heterostructures.

\end{abstract}

\maketitle

High-electron-mobility transistors (HEMTs) based on AlGaN/GaN have emerged as key components for high-frequency communication systems, ranging from satellite links to terrestrial networks. Despite the large lattice mismatch with group-III nitrides and the low thermal conductivity of Si, Si remains an attractive substrate for HEMTs when mass production and fabrication cost are key considerations and operation at high temperatures or very high power levels is not required. Furthermore, CMOS-compatible HEMTs enable integration with Si-based drivers and digital signal processors,\cite{then202550} opening the door to more compact and efficient next-generation communication systems. Particular interest lies in N-polar nitride HEMTs with AlN back barriers on Si(111), owing to the advantages in layer design enabled by N-polarity.\cite{romanczyk2017demonstration} Such devices have demonstrated record output power densities of up to 8.0~W/mm at 94~GHz,\cite{romanczyk2017demonstration,romanczyk2021evaluation} highlighting their strong potential for high-frequency applications. 

AlN films on Si have been extensively investigated as buffer layers and templates for III-nitride heterostructures, including metal-polar HEMTs on Si.\cite{cheng2008algan,frayssinet2017influence,ganguly2014plasma,cordier2018influence} In contrast, demonstrations of N-polar nitride device heterostructures directly grown on Si remain comparatively limited. Recently, N-polar heterostructures on Si have been demonstrated using NbN interlayers to facilitate polarity control.\cite{barbier2026n} While this approach provides a route to N-polarity, the introduction of a conductive interlayer is undesirable for device structures requiring high electrical isolation. This motivates the growth of N-polar AlN directly on Si without a conductive polarity-control interlayer while maintaining high electrical isolation and structural quality.

Previous studies have demonstrated that the polarity and structural properties of AlN grown on Si(111) are strongly influenced by the nucleation and growth conditions.\cite{fan2024polarity,roshko2020eutectic,wang2022interfacial} However, how the growth conditions governing polarity selection simultaneously affect the electrical isolation and structural quality of AlN/Si(111) remains insufficiently understood. Here, we systematically investigate the effects of substrate temperature and Al flux on these coupled properties. We show that Al-rich conditions promote N-polarity and improved structural quality but simultaneously enhance unwanted conductivity, revealing a growth-induced trade-off among polarity, electrical isolation, and structural quality.




75-nm-thick AlN films were grown on 2-inch high-resistivity ($>20$~k$\Omega$·cm; sheet resistance \rsh$>667$~k$\Omega$/sq) Si(111) substrates using plasma-assisted molecular beam epitaxy (MBE). About 100~nm-thick thermal oxide on as-received Si substrates was first removed by dipping in 5$\%$ HF for 6 minutes followed by deionized-water rinse. Thermal deoxidation of the Si substrates was performed in the growth chamber and confirmed by the appearance of the Si(111)$-(7\times7)$ surface reconstruction. The reported surface temperature of 830\celsius for this reconstruction\cite{telieps19857} was used as a reference to calibrate the pyrometer of the MBE system. All substrate temperatures reported in this work refer to the pyrometer readings. Active N species were supplied using an RF plasma power of 400~W and an N$_{2}$ gas flow rate of 3.5~sccm. These conditions resulted in a growth rate of 4.5~nm/min for \textit{c}-plane AlN under Al-rich conditions, corresponding to an N flux of \fn$=4.5$~nm/min.

The substrate temperature \ts and Al flux \fal were used as the MBE growth variables in this study. \ts was set to 575, 675, and 775\celsius, while \fal was varied from 3.5 to 8.5~nm/min. Prior to AlN growth, 1~monolayer (ML) of Al was deposited at the growth temperature to induce the Al/Si(111) \(\gamma\) phase,\cite{vg2005atomic,groger2001ultrathin,yasutake1998molecular,dasgupta2009growth} in which the dangling bonds of the Si surface atoms are fully saturated by Al, thereby preventing the formation of unintentional amorphous SiN$_{x}$. Such an Al ML can also lead to a smaller lattice mismatch with AlN compared with the case of direct AlN growth on Si.\cite{shin2018epitaxial} Subsequently, Al and N fluxes were supplied simultaneously and continuously for AlN growth. In situ reflection high-energy electron diffraction (RHEED) and 30\% KOH etching followed by scanning electron microscopy (SEM) imaging were employed to determine the polarity of the grown AlN films. The electrical properties of the samples were assessed by Hall-effect measurements using the van der Pauw method with In dots as ohmic contacts at room temperature. The structural quality of the films was evaluated by x-ray diffraction (XRD) $\omega$ scans (rocking curves) and bright-field scanning transmission electron microscopy (BF-STEM). Energy-dispersive x-ray spectroscopy (EDX) was employed to investigate the elemental distribution within the samples. The surface morphology of samples grown under N-rich and Al-rich conditions was investigated using atomic force microscopy (AFM) in tapping mode.


\begin{figure}[t!]
\centering
\includegraphics*[width=8.0cm]{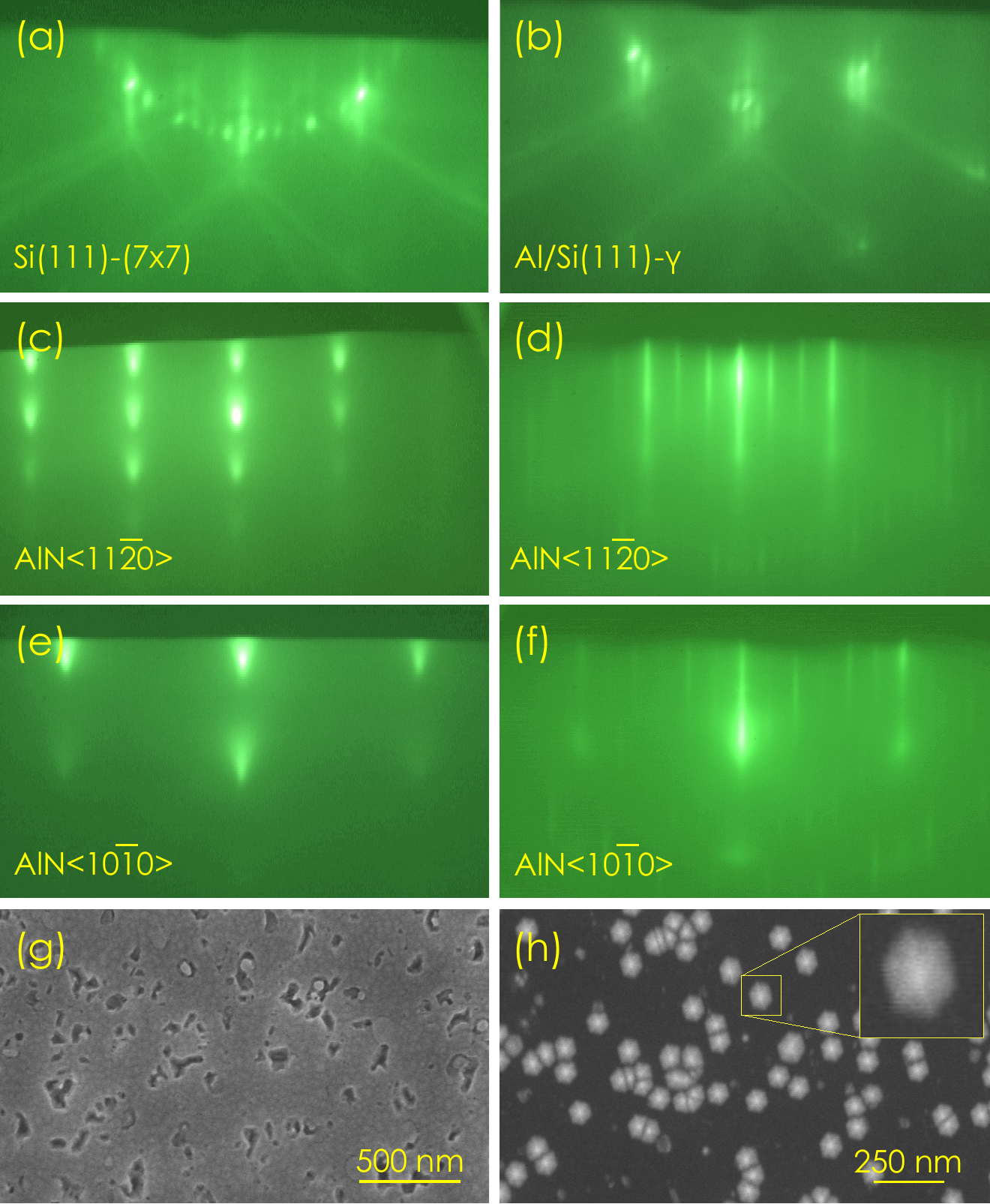} 
\caption{[(a)--(f)] Representative RHEED patterns of (a) Si(111) before growth, (b) after deposition of 1~ML of Al, and [(c)--(f)] after subsequent growth of AlN under [(c), (e)] N-rich (Al flux: 3.5~nm/min) and [(d), (f)] Al-rich (Al flux: 8.5~nm/min) conditions. [(g), (h)] SEM images obtained after KOH etching of the AlN layers grown under (g) N-rich and (h) Al-rich conditions. The RHEED patterns were recorded along the [(a)--(d)] Si $\langle 1\bar{1}0\rangle$ and [(e), (f)] Si $\langle 11\bar{2}\rangle$ zone axes at [(a), (b)] 775\celsius and [(c)--(f)] below 300\celsius. The substrate temperature during AlN growth was 775\celsius.}
\label{fig_rheed}
\end{figure}
An oxide-free Si(111) surface was first confirmed by the appearance of the $(7\times7)$ reconstruction [Fig.~\ref{fig_rheed}(a)]. Following Al deposition, the RHEED pattern changed to a $(1\times1)$ pattern with pronounced satellite peaks [Fig.~\ref{fig_rheed}(b)]. These satellite peaks are attributed to an Al ML with an in-plane lattice constant $10/9$ times that of Si, corresponding to the Al/Si(111)$-\gamma$ phase.\cite{yasutake1998molecular} 

As commonly observed in nitride MBE growth,\cite{heying2000optimization,cho2026dependence} N-rich conditions resulted in spotty RHEED patterns [Fig.~\ref{fig_rheed}(c) and (e)], whereas Al-rich conditions resulted in streaky RHEED patterns [Fig.~\ref{fig_rheed}(d) and (f)]. However, the samples grown at 575\celsius exhibited spotty RHEED patterns even under Al-rich conditions, suggesting that the relatively low growth temperature suppresses adatom surface diffusion. The epitaxial relations Si$\langle 1\bar{1}0\rangle$$||$AlN$\langle 11\bar{2}0\rangle$ and Si$\langle 11\bar{2}\rangle$$||$AlN$\langle 10\bar{1}0\rangle$ are also observed. [Figs.~\ref{fig_rheed}(a)--\ref{fig_rheed}(f)] More importantly, the samples grown under Al-rich conditions exhibited clear surface reconstruction features, with a superposition of $(3\times3)$ and $(6\times6)$ reconstructions, characteristic of N-polar AlN surfaces.\cite{dasgupta2012growth,ledyaev2014n} The polarity of the AlN films grown on Si(111) was further examined by ex situ KOH etching. Samples grown under N-rich conditions exhibited rough surfaces without considerable KOH-induced etching [Fig.~\ref{fig_rheed}(g)], whereas the Al-rich samples developed hexagonal pyramids [Fig.~\ref{fig_rheed}(h) and the inset], a characteristic morphology typically observed for N-polar nitrides after KOH etching.\cite{zhang2022molecular,pampili2024geometrically} Interestingly, the Al/Si(111)$-\gamma$ phase, which is expected to facilitate the nucleation of N-polar AlN,\cite{dasgupta2009growth} does not determine the polarity of the overgrown AlN. Instead, under the growth protocol and parameter range investigated here, N-polarity is obtained only when nucleation proceeds under Al-rich conditions (\fal$>$\fn).

\begin{figure}[t!]
\centering
\includegraphics*[width=8.0cm]{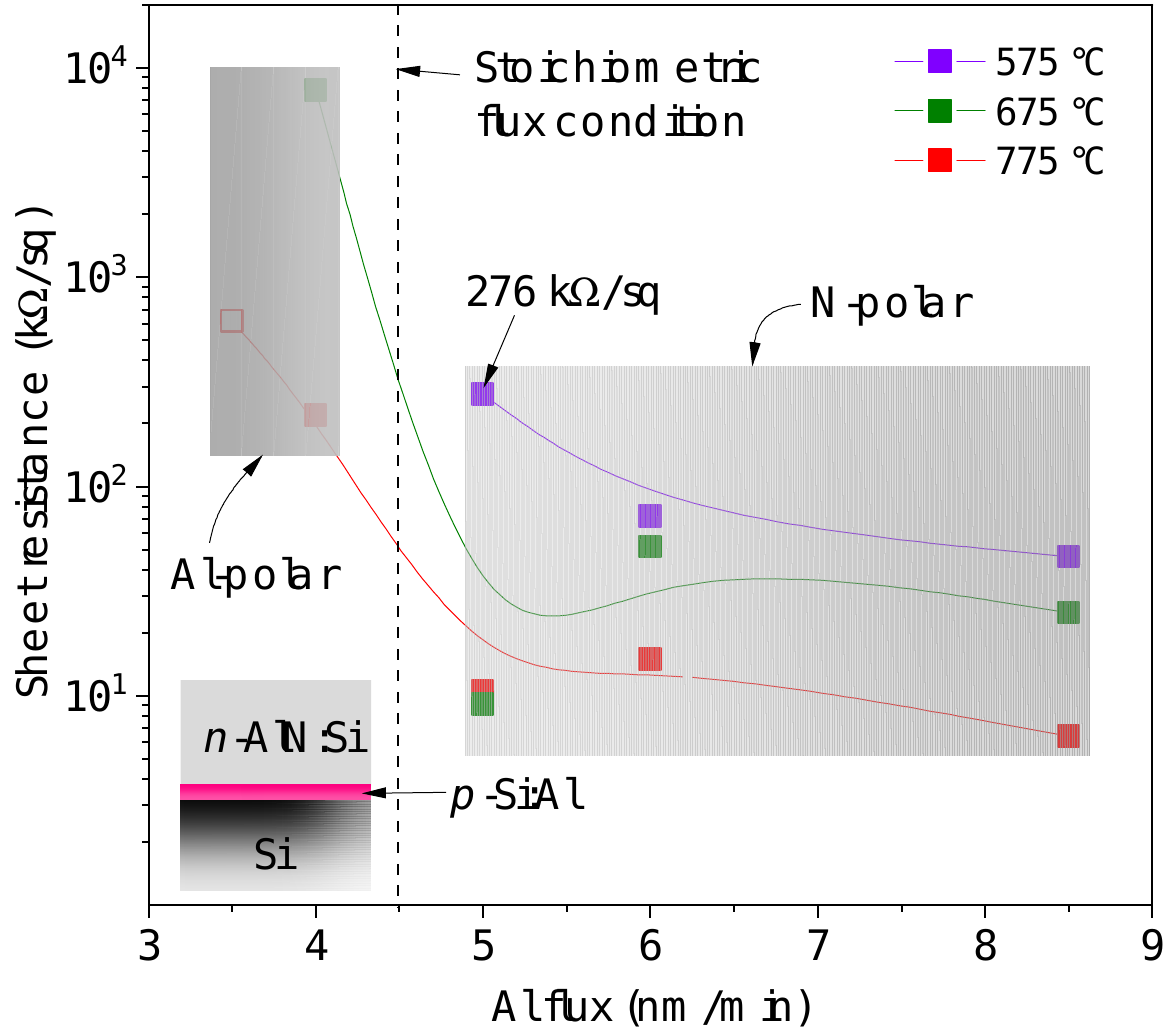} 
\caption{Room-temperature sheet resistance of AlN/Si(111) stacks grown at different substrate temperatures and Al fluxes. The lines are guides to the eye. All the samples exhibit \textit{p}-type conductivity (filled squares), except for the sample grown at 775\celsius with an Al flux of 3.5~nm/min, which exhibits \textit{n}-type conductivity (open square). The dashed line indicates the stoichiometric flux condition (\fal$=$\fn), which separates the two polarities of the AlN films. The inset schematically illustrates the regions of \textit{p}- and \textit{n}-type conductivity associated with Al in-diffusion and Si out-diffusion, respectively.}
\label{fig_hall}
\end{figure}

\begin{figure}[t!]
\centering
\includegraphics*[width=8.0cm]{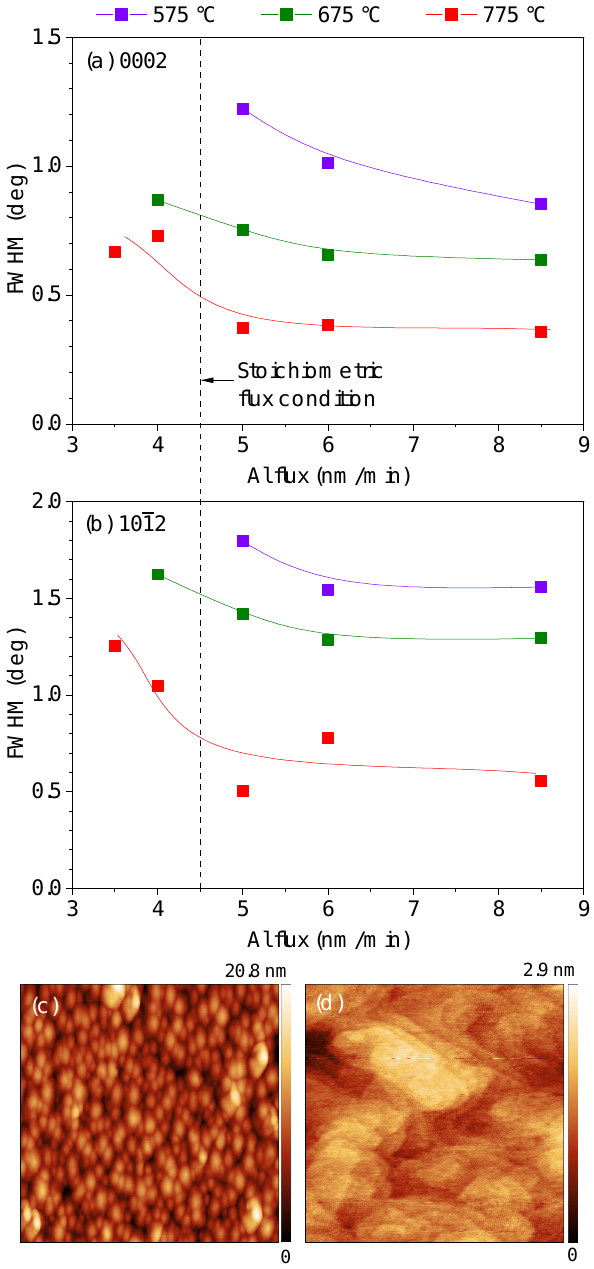} 
\caption{[(a), (b)] Full width at half maximum (FWHM) of (a) 0002 and (b) $10\bar{1}2$ XRD $\omega$ scans for AlN films grown on high-resistivity Si(111) substrates at different growth temperatures and Al fluxes. The lines are guides to the eye. [(c), (d)] $1\times1$~$\mu$m$^{2}$ AFM micrographs of AlN films grown at 775\celsius using Al fluxes of (c) 3.5~nm/min and (d) 8.5~nm/min. The root-mean-square (RMS) roughness values of the surfaces in (c) and (d) are 2.5 and 0.3~nm, respectively.}
\label{fig_xrd}
\end{figure}

Figure~\ref{fig_hall} shows the sheet resistance \rsh of AlN/Si stacks grown at different substrate temperatures \ts and Al fluxes \fal. The measured \rsh spans more than four orders of magnitude, ranging from approximately 5 k$\Omega$/sq to values exceeding the measurement range. Samples grown at lower \fal and lower \ts tend to exhibit higher \rsh. This trend becomes particularly pronounced upon crossing the stoichiometric flux condition into the N-rich regime, with \rsh increasing further as the growth conditions become increasingly N-rich. (The samples grown at 575\celsius under N-rich conditions could not be measured because their resistance exceeded the measurement range.) Most samples exhibit \textit{p}-type conduction, which is consistent with Al in-diffusion into the Si substrate and the resulting acceptor doping reported previously for AlN/Si interfaces.\cite{wei2020diffusion,dasgupta2012growth} Conversely, Si can out-diffuse into the overgrown AlN and act as an \textit{n}-type dopant in AlN.\cite{dasgupta2012growth,ganguly2014plasma} However, because of the large difference in bandgap energy between AlN and Si, conduction through \textit{p}-type Si is expected to dominate the apparent electrical conduction of the AlN/Si structure. Al acceptor doping in Si is likely enhanced by both increasing \ts and \textit{increasing} \fal,  while Si donor doping is likely to be enhanced with increasing \ts and \textit{decreasing} \fal, as Si donors occupy the cation Al sites in AlN. Indeed, considerable Si out-diffusion has been reported for AlN grown on Si by MBE under N-rich conditions.\cite{ganguly2014plasma} The apparent \textit{n}-type conduction of the sample grown at \fal$=$3.5~nm/min and 775\celsius may therefore indicate that Si donor doping promoted by the high \ts and the lowest \fal becomes dominant over Al acceptor doping suppressed by the sufficiently low \fal. The AlN sample grown at 575\celsius under slightly Al-rich conditions (\fal$=5$~nm/min; \fal/\fn = 1.11) exhibits an \rsh of 276~k$\Omega$/sq and N-polarity.  

\begin{figure*}[t!]
\centering
\includegraphics*[width=13.5cm]{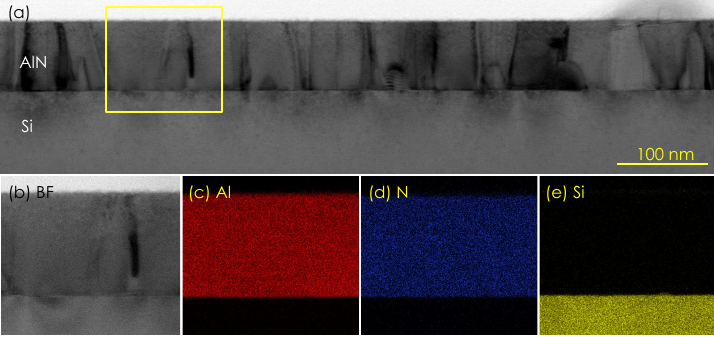} 
\caption{(a) Cross-sectional BF-STEM micrograph of an AlN film grown on high-resistivity Si(111) at 775\celsius with an Al flux of 8.5~nm/min. (b) Higher-magnification BF-STEM micrograph of the region indicated by the yellow box in (a), along with the corresponding EDX elemental maps of (c) Al, (d) N, and (e) Si.}
\label{fig_tem}
\end{figure*}

We now turn to the structural quality of AlN films grown on Si(111). Figures~\ref{fig_xrd}(a) and (b) show the full width at half maximum (FWHM) of XRD $\omega$ scans in symmetric geometry for the on-axis 0002 reflection and in skew-symmetric geometry for the $10\bar{1}2$ reflection of AlN films grown on Si(111) at different \ts and \fal. It is observed that the structural quality of the AlN films improves with increasing \ts and under Al-rich conditions compared to N-rich conditions, as evidenced by the smaller FWHM values. This improvement in structural properties can be attributed to an enhanced adatom diffusion length under Al-rich conditions, which is further evidenced by the absence and presence of clear atomic steps on the surface in N-rich and Al-rich conditions, respectively [Figs.~\ref{fig_xrd}(c) and \ref{fig_xrd}(d)]. 

In heterovalent epitaxy, chemical reactions at the interface between adjacent materials can be a major concern and may play an important role in determining the structural properties of the system.\cite{cho2012structural,cho2012effects} The sample grown at the highest \ts and \fal shows no evidence of widespread interfacial reactions [Figs.~\ref{fig_tem}(a)--\ref{fig_tem}(e)]. Approximately 10 nm tails are observed in the Al and Si EDX line scans across the interface (not shown here). However, such broadening can arise from the limited spatial resolution of EDX even for sharp interfaces. While low-level interdiffusion at the AlN/Si interface remains below the detection limit of EDX, the conductivity behavior observed in Fig.~\ref{fig_hall} is consistent with prior secondary ion mass spectrometry (SIMS) studies reporting Al in-diffusion into Si (Refs.~\onlinecite{wei2020diffusion,dasgupta2012growth}) and Si out-diffusion into AlN (Refs.~\onlinecite{ganguly2014plasma,dasgupta2012growth}). Together with the high density of dislocations observed in Fig.~\ref{fig_tem}(a), these results suggest that the broadening of the rocking curves in Figs.~\ref{fig_xrd}(a) and \ref{fig_xrd}(b) is predominantly associated with dislocations arising from the large lattice mismatch, rather than with widespread interfacial reactions. However, evidence of chemical reactions was found in a few localized regions in samples grown under highly Al-rich conditions; these reactions will be discussed in a separate work.

In summary, for N-polar AlN/Si intended for HEMT structures, the results reveal a trade-off among polarity, electrical isolation, and structural quality. Al-rich growth conditions promote N-polarity and improve structural quality but also increase unwanted conductivity, particularly at higher \ts. The observed conductivity trend is consistent with Al acceptor incorporation into Si reported previously.\cite{dasgupta2012growth,wei2020diffusion} By reducing \ts while maintaining slightly Al-rich conditions, an N-polar AlN/Si stack with a sheet resistance \rsh of 276~k$\Omega$/sq was achieved, more than one order of magnitude higher than that of samples grown at \ts$=775$\celsius, although still lower than the Si substrate value ($\geq 667$~k$\Omega$/sq). These findings suggest that further optimization of electrical isolation and structural quality may be achieved through multistep growth schemes involving controlled variations in flux conditions and substrate temperature.

The authors thank Achim~Trampert for helpful discussions, Oliver~Bierwagen for reviewing the manuscript, and Hans-Peter~Schönherr, Claudia~Herrmann and Jingxuan~Kang for maintaining the MBE system. Nhu Tran and Roman Gröger acknowledge financial support for this research from the Czech Science Foundation grant No. 24-12526S.

\section{Author Declarations}
The authors have no conflicts to disclose.

\section{Data Availability}
The data that support the findings of this study are available from the corresponding authors upon reasonable request.

\bibliographystyle{aipnum4-1}
\bibliography{references}

@article{ledyaev2014n,
  title={N-polar AlN thin layers grown on Si (111) by plasma-assisted MBE},
  author={Ledyaev, Oleg and Pandikunta, Mahesh and Nikishin, Sergey},
  journal={Japanese Journal of Applied Physics},
  volume={53},
  number={5},
  pages={050306},
  year={2014},
  publisher={The Japan Society of Applied Physics}
}

@article{dasgupta2009growth,
  title={Growth of high quality N-polar AlN (0001) on Si (111) by plasma assisted molecular beam epitaxy},
  author={Dasgupta, Sansaptak and Wu, F and Speck, JS and Mishra, UK},
  journal={Applied Physics Letters},
  volume={94},
  number={15},
  pages={151906},
  year={2009},
  publisher={AIP Publishing}
}

@article{yasutake1998molecular,
  title={Molecular beam epitaxial growth of AlN single crystalline films on Si (111) using radio-frequency plasma assisted nitrogen radical source},
  author={Yasutake, K and Takeuchi, A and Kakiuchi, H and Yoshii, K},
  journal={Journal of Vacuum Science \& Technology A},
  volume={16},
  number={4},
  pages={2140--2147},
  year={1998},
  publisher={American Vacuum Society}
}

@article{vg2005atomic,
  title={Atomic structure of the Al/Si (111) phases studied using STM and total-energy calculations},
  author={Kotlyar, V. and Saranin, A. and Zotov, A. and  Kasyanova, T. and Chukurov, E. and Pisarenko, I. and  Lifshits, V.},
  journal={e-Journal of Surface Science and Nanotechnology},
  volume={3},
  pages={55--62},
  year={2005},
  publisher={The Japan Society of Vacuum and Surface Science}
}

@article{groger2001ultrathin,
  title={Ultrathin Al layers on Si (111) and Si (100): structures and phase transitions},
  author={Gr{\"o}ger, RM and Barczewski, MR},
  journal={Surface and Interface Analysis},
  volume={32},
  number={1},
  pages={154--160},
  year={2001},
  publisher={Wiley Online Library}
}

@article{wei2020diffusion,
  title={Al diffusion at AlN/Si interface and its suppression through substrate nitridation},
  author={Wei, Lai and Yang, Xuelin and Shen, Jianfei and Liu, Danshuo and Cai, Zidong and Ma, Cheng and He, Xiaoguang and Tang, Jun and Qi, Shengli and Xu, Fujun and others},
  journal={Applied Physics Letters},
  volume={116},
  number={23},
  pages={232105},
  year={2020},
  publisher={AIP Publishing}
}

@article{zhang2022molecular,
  title={Molecular beam homoepitaxy of N-polar AlN: Enabling role of aluminum-assisted surface cleaning},
  author={Zhang, Zexuan and Hayashi, Yusuke and Tohei, Tetsuya and Sakai, Akira and Protasenko, Vladimir and Singhal, Jashan and Miyake, Hideto and Xing, Huili Grace and Jena, Debdeep and Cho, YongJin},
  journal={Science Advances},
  volume={8},
  number={36},
  pages={eabo6408},
  year={2022},
  publisher={American Association for the Advancement of Science}
}

@article{pampili2024geometrically,
  title={Geometrically controlled microscale patterning and epitaxial lateral overgrowth of nitrogen-polar GaN},
  author={Pampili, Pietro and Zubialevich, Vitaly Z and Parbrook, Peter J},
  journal={Crystal Growth \& Design},
  volume={24},
  number={15},
  pages={6095},
  year={2024}
}

@article{heying2000optimization,
  title={Optimization of the surface morphologies and electron mobilities in GaN grown by plasma-assisted molecular beam epitaxy},
  author={Heying, B and Smorchkova, I and Poblenz, C and Elsass, C and Fini, P and Den Baars, S and Mishra, U and Speck, JS},
  journal={Applied Physics Letters},
  volume={77},
  number={18},
  pages={2885--2887},
  year={2000},
  publisher={American Institute of Physics}
}

@article{cho2026dependence,
  title={Dependence of the Mn sticking coefficient on Ga-rich, N-rich, and Ga/N-flux-free conditions in GaN grown by plasma-assisted molecular beam epitaxy},
  author={Cho, YongJin and Yu, Changkai and Grace Xing, Huili and Jena, Debdeep},
  journal={Journal of Vacuum Science \& Technology A},
  volume={44},
  number={2},
  pages={022706},
  year={2026},
  publisher={AIP Publishing}
}

@article{cho2012structural,
  title={Structural properties of InN films grown on O-face ZnO (0001) by plasma-assisted molecular beam epitaxy},
  author={Cho, YongJin and Brandt, Oliver and Korytov, Maxim and Albrecht, Martin and Kaganer, Vladimir M and Ramsteiner, Manfred and Riechert, Henning},
  journal={Applied Physics Letters},
  volume={100},
  number={15},
  pages={152105},
  year={2012},
  publisher={AIP Publishing}
}

@article{ganguly2014plasma,
  title={Plasma MBE growth conditions of AlGaN/GaN high-electron-mobility transistors on silicon and their device characteristics with epitaxially regrown ohmic contacts},
  author={Ganguly, Satyaki and Verma, Jai and Xing, Huili and Jena, Debdeep},
  journal={Applied Physics Express},
  volume={7},
  number={10},
  pages={105501},
  year={2014},
  publisher={The Japan Society of Applied Physics}
}

@article{then202550,
  title={50 nm DrGaN in 3D monolithic GaN MOSHEMT and Silicon PMOS process on 300 mm GaN-on-Si (111)},
  author={Then, Han Wui and Radosavljevic, M and Bader, S and Zubair, A and Vora, H and Koirala, P and Beumer, M and Nordeen, P and Vyatskikh, A and Hoff, T and others},
  journal={Power Electronic Devices and Components},
  volume={10},
  pages={100074},
  year={2025},
  publisher={Elsevier}
}

@article{romanczyk2017demonstration,
  title={Demonstration of constant 8 W/mm power density at 10, 30, and 94 GHz in state-of-the-art millimeter-wave N-polar GaN MISHEMTs},
  author={Romanczyk, Brian and Wienecke, Steven and Guidry, Matthew and Li, Haoran and Ahmadi, Elaheh and Zheng, Xun and Keller, Stacia and Mishra, Umesh K},
  journal={IEEE Transactions on Electron Devices},
  volume={65},
  number={1},
  pages={45--50},
  year={2017},
  publisher={IEEE}
}

@article{romanczyk2021evaluation,
  title={Evaluation of linearity at 30 GHz for N-polar GaN deep recess transistors with 10.3 W/mm of output power and 47.4\% PAE},
  author={Romanczyk, Brian and Guidry, Matthew and Zheng, Xun and Shrestha, Pawana and Li, Haoran and Ahmadi, Elaheh and Keller, Stacia and Mishra, Umesh K},
  journal={Applied Physics Letters},
  volume={119},
  number={7},
  pages={072105},
  year={2021},
  publisher={AIP Publishing}
}

@article{cho2012effects,
  title={Effects of Ga on the growth of InN on O-face ZnO (0001) by plasma-assisted molecular beam epitaxy},
  author={Cho, YongJin and Korytov, Maxim and Albrecht, Martin and Riechert, Henning and Brandt, Oliver},
  journal={Applied Physics Letters},
  volume={101},
  number={5},
  pages={052103},
  year={2012},
  publisher={AIP Publishing}
}

@article{telieps19857,
  title={The (7$\times$ 7)↔(1$\times$ 1) phase transition on Si (111)},
  author={Telieps, W and Bauer, E},
  journal={Surface Science},
  volume={162},
  number={1-3},
  pages={163--168},
  year={1985},
  publisher={Elsevier}
}

@article{barbier2026n,
  title={N-polar AlGaN HEMTs on silicon for reduced contact resistance},
  author={Barbier-Cueil, A and Mehta, J and P{\'e}d{\`e}ches, A and Medjdoub, F and Semond, F},
  journal={Applied Physics Letters},
  volume={128},
  number={26},
  pages={262103},
  year={2026},
  publisher={AIP Publishing}
}

@article{cheng2008algan,
  title={AlGaN/GaN high electron mobility transistors grown on 150 mm Si (111) substrates with high uniformity},
  author={Cheng, Kai and Leys, Maarten and Degroote, Stefan and Derluyn, Joff and Sijmus, Brian and Favia, Paola and Richard, Olivier and Bender, Hugo and Germain, Marianne and Borghs, Gustaaf},
  journal={Japanese Journal of Applied Physics},
  volume={47},
  number={3R},
  pages={1553},
  year={2008}
}

@article{frayssinet2017influence,
  title={Influence of metal-organic vapor phase epitaxy parameters and Si (111) substrate type on the properties of AlGaN/GaN HEMTs with thin simple buffer},
  author={Frayssinet, Eric and Leclaire, Paul and Mohdad, Jad and Latrach, Soumaya and Chenot, S{\'e}bastien and Nemoz, Maud and Damilano, Benjamin and Cordier, Yvon},
  journal={Physica Status Solidi A},
  volume={214},
  number={4},
  pages={1600419},
  year={2017},
  publisher={Wiley Online Library}
}

@article{fan2024polarity,
  title={Polarity control and crystalline quality improvement of AlN thin films grown on Si (111) substrates by molecular beam epitaxy},
  author={Fan, Shizhao and Yin, Yuhao and Liu, Rong and Zhao, Haiyang and Liu, Zhenghui and Sun, Qian and Yang, Hui},
  journal={Journal of Applied Physics},
  volume={136},
  number={14},
  pages={145301},
  year={2024},
  publisher={AIP Publishing}
}

@article{roshko2020eutectic,
  title={Eutectic formation, V/III ratio, and controlled polarity inversion in nitrides on silicon},
  author={Roshko, Alexana and Brubaker, Matt D and Blanchard, Paul T and Harvey, Todd E and Bertness, Kris A},
  journal={Physica Status Solidi B},
  volume={257},
  number={4},
  pages={1900611},
  year={2020},
  publisher={Wiley Online Library}
}

@article{wang2022interfacial,
  title={Interfacial modulated lattice-polarity-controlled epitaxy of III-nitride heterostructures on Si (111)},
  author={Wang, Ping and Wang, Ding and Mondal, Shubham and Wu, Yuanpeng and Ma, Tao and Mi, Zetian},
  journal={ACS Applied Materials \& Interfaces},
  volume={14},
  number={13},
  pages={15747--15755},
  year={2022},
  publisher={ACS Publications}
}

@article{cordier2018influence,
  title={Influence of AlN growth temperature on the electrical properties of buffer layers for GaN HEMTs on silicon},
  author={Cordier, Yvon and Comyn, R{\'e}mi and Frayssinet, Eric and Khoury, Mario and Lesecq, Marie and Defrance, Nicolas and De Jaeger, Jean-Claude},
  journal={Physica Status Solidi A},
  volume={215},
  number={9},
  pages={1700637},
  year={2018},
  publisher={Wiley Online Library}
}

@article{shin2018epitaxial,
  title={Epitaxial growth of single-crystalline AlN layer on Si (111) by DC magnetron sputtering at room temperature},
  author={Shin, In-Su and Kim, Jongmyeong and Lee, Donghyun and Kim, Donghyun and Park, Yongjo and Yoon, Euijoon},
  journal={Japanese Journal of Applied Physics},
  volume={57},
  number={6},
  pages={060306},
  year={2018},
  publisher={The Japan Society of Applied Physics}
}

@article{dasgupta2012growth,
  title={Growth and characterization of N-polar GaN films on Si (111) by plasma assisted molecular beam epitaxy},
  author={Dasgupta, Sansaptak and Wu, Nidhi and Speck, James S and Mishra, Umesh K},
  journal={Japanese Journal of Applied Physics},
  volume={51},
  number={11R},
  pages={115503},
  year={2012}
}
\end{document}